\documentclass[twocolumn,a4paper,10pt]{article}

\usepackage[utf8]{inputenc}
\usepackage[T1]{fontenc}
\usepackage{lmodern}
\usepackage[english]{babel}
\usepackage[margin=2.0cm]{geometry}
\usepackage{graphicx}
\usepackage{booktabs}
\usepackage{microtype}
\usepackage{xcolor}
\usepackage{hyperref}
\usepackage{xurl}
\usepackage{caption}
\usepackage{subcaption}
\usepackage{float}
\usepackage[round,authoryear]{natbib}

\hypersetup{
  colorlinks=true,
  linkcolor=black,
  citecolor=blue!50!black,
  urlcolor=blue!50!black,
}

\title{\textbf{Eleven Years of BRACIS:\\
A Meta-Scientific Study of the Brazilian Conference on Intelligent Systems}}

\author{%
\parbox{\linewidth}{\centering
\textbf{Thales Sales Almeida}\textsuperscript{1,2} \quad
\textbf{Giovana Kerche Bonás}\textsuperscript{1,2} \quad
\textbf{Thiago Laitz}\textsuperscript{1,2} \\[0.6ex]
\textbf{João Guilherme Alves Santos}\textsuperscript{1,2} \quad
\textbf{Hugo Abonizio}\textsuperscript{2} \quad
\textbf{Roseval Malaquias Junior}\textsuperscript{2} \\[0.6ex]
\textbf{Marcos Piau}\textsuperscript{2} \quad
\textbf{Celio Larcher}\textsuperscript{2} \quad
\textbf{Ramon Pires}\textsuperscript{2} \quad
\textbf{Rodrigo Nogueira}\textsuperscript{2} \\[1.2ex]
\normalfont\normalsize
\textsuperscript{1}Tropic AI \qquad \textsuperscript{2}Maritaca AI}
}

\begin{document}
\twocolumn[
  \begin{@twocolumnfalse}
    \maketitle
    \begin{abstract}
    \noindent The Brazilian Conference on Intelligent Systems (BRACIS) is the main national venue for Artificial Intelligence research in Brazil, hosted by the Brazilian Computer Society since 2012 and publishing work from institutions across the country. Across eleven years, from 2015 to 2025, we build a per-paper record of all 1,066 accepted papers from DBLP metadata, 6,765 Google Scholar citations, and the paper full texts, and use it to ask what BRACIS publishes, who publishes it, and which work gets cited. Large Language Model research grows from zero before 2020 to 19\% of papers in 2024, on top of a base of Machine Learning, Computer Vision, and Optimization work. The community is hourglass-shaped: 80.5\% of 2,623 authors appear in a single edition, while institutions return at nearly three times the author rate. Citations are heavily concentrated, with the top 1\% of papers carrying 27\% of the total. Openness practices have grown, with artifact release rising from 8.9\% of papers in 2015 to 57.3\% in 2023, and we find a notable correlation between having an arXiv preprint and higher citation counts. Since proceedings sit behind IEEE and Springer paywalls and only 7.4\% of papers have a preprint, most BRACIS work is hard to reach for readers without institutional access.
    \end{abstract}
    \vspace{1.0em}
  \end{@twocolumnfalse}
]

\section{Introduction}\label{sec:intro}

As Artificial Intelligence (AI) research grows in scale and specialization, regional conferences play a fundamental part in shaping national priorities and recognizing work
that does not fit the agendas or interests of the largest international venues. The Brazilian Conference on Intelligent Systems (BRACIS) is one such conference. Hosted by the Brazilian Computer Society since 2012, BRACIS published 1,066 papers between 2015 and 2025 covering topics from metaheuristics to LLMs.


Several recent papers apply bibliometric methods to large international venues \citep{mohammad2020state,pramanick2025nature,aclcrown2025}, and a parallel Brazilian research has done the same for some SBC conferences \citep{limafilho2023csbcset,carvalho2024sbsi,carvalho2024wcama,procopio2017sbbd,mendonca2022sbes,nunes2026propor}. BRACIS was briefly studied by \citet{albuquerque2024nlp}, which restricts itself to NLP-and-social-media papers across BRACIS and other Brazilian events. In this study we aim to brodly explore the research around BRACIS.

\begin{figure*}
    \centering
    \includegraphics[width=0.7\linewidth]{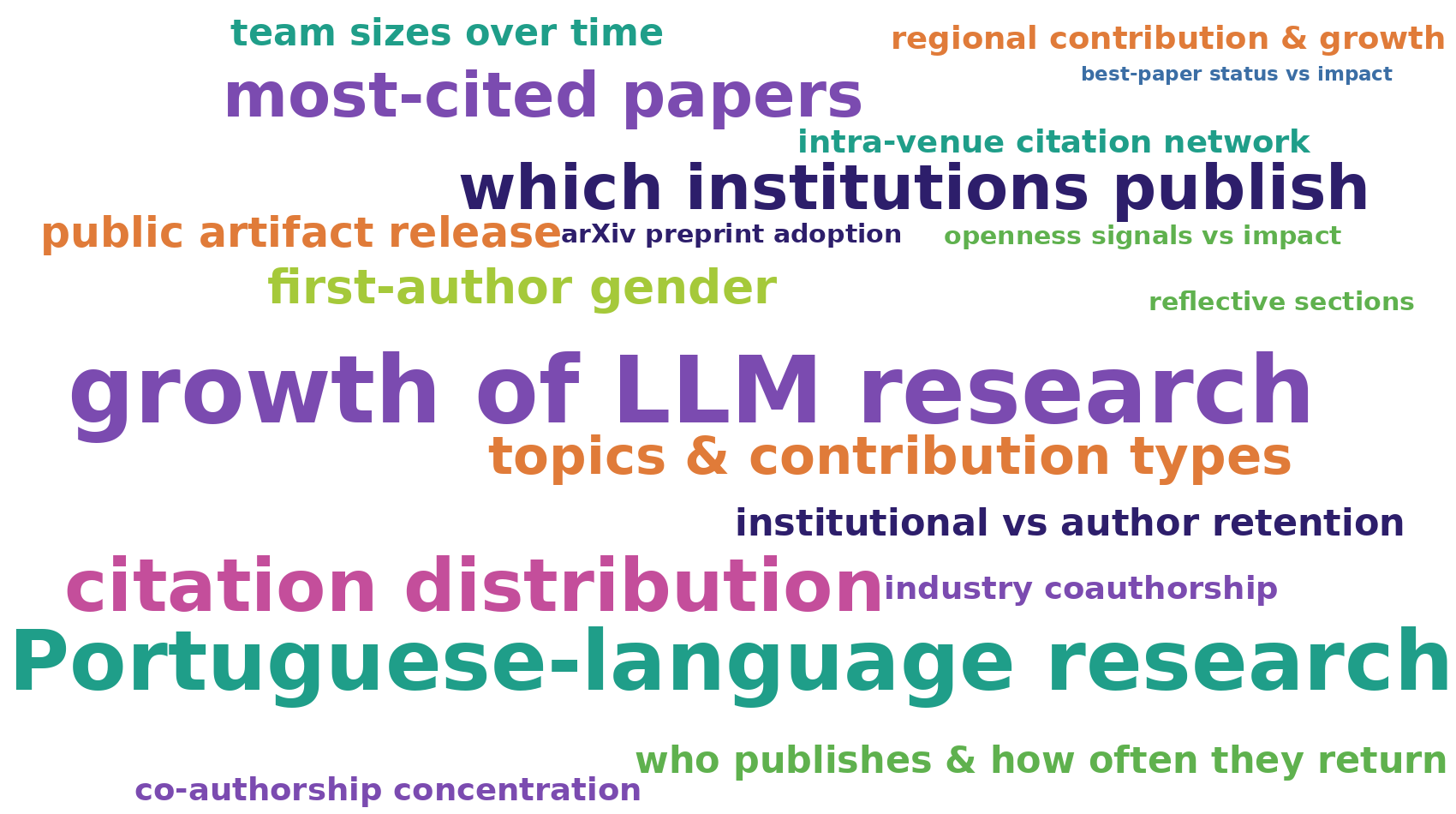}
      \caption{Wordcloud of topics investigated in this paper.}
  \label{fig:phrasecloud}

\end{figure*}

For our study, we built a per-paper record of all 1{,}066 papers published in BRACIS in the last 11 years. We explore multiple aspects of the venue, including its topical and contribution-type composition, the authors and institutions behind it, the distribution and concentration of citations, the intra-venue citation network, and the adoption of openness practices such as artifact release, arXiv posting, and industry collaboration. Figure~\ref{fig:phrasecloud} shows a visual representation of topics explored in this paper.

We organize the analysis around the three-pillar layout, similarly to \citet{nunes2026propor}'s: thematic landscape, community structure, and scientific impact.
\begin{itemize}
  \item \emph{Thematic landscape: which topics and contribution types define BRACIS, and how has the topical composition shifted across the decade?}
  \item \emph{Community: who publishes at BRACIS, how do authors and institutions interact, and what does retention look like at each level?}
  \item \emph{Scientific impact: how are citations distributed across BRACIS papers, what predicts impact, and how do openness signals correlate with it?}
\end{itemize}

\begin{figure}[t]
  \centering
  \includegraphics[width=\linewidth]{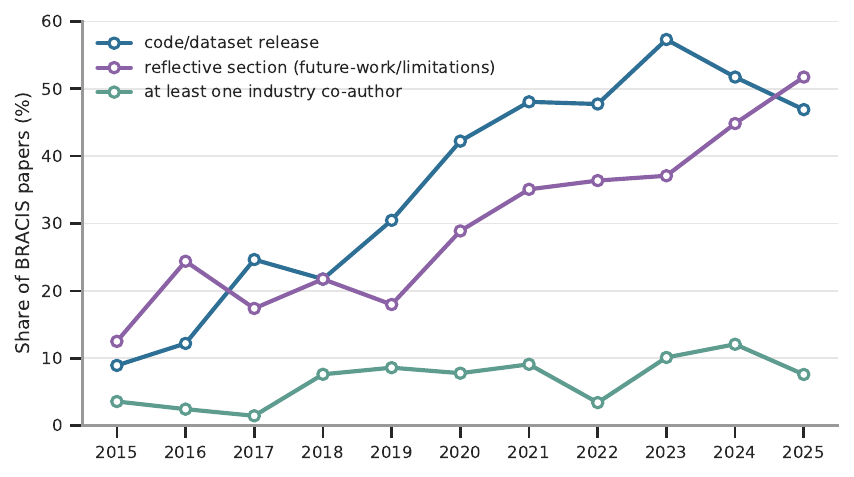}
  \caption{Three longitudinal trends at BRACIS, 2015 to 2025. Each line shows the share of accepted papers each year exhibiting the property: a public code or data release URL, at least one industry-affiliated co-author, or an explicit \emph{Limitations}, \emph{Future Work}, or \emph{Conclusion and Future Work} section.}
  \label{fig:teaser}
\end{figure}

\section{Related Work}\label{sec:related}

\paragraph{Bibliometric studies of AI venues.}
Large-scale bibliometric analyses of AI publication venues have become an established line of meta-scientific research. \citet{mohammad2020state} traces productivity, research focus, impact, and gender representation across the entire ACL Anthology, with a companion study dedicated to citation patterns \citep{mohammad2020citations}. More recently, \citet{pramanick2025nature} propose a taxonomy of contribution types for NLP papers and apply it at ACL Anthology scale, an approach methodologically parallel to our LLM-based contribution classification. On the data side, \citet{wahle2022d3} release \textsc{D3}, a DBLP-derived dataset that underpins several recent multi-venue analyses, including \citet{aclcrown2025}, while \citet{cardoso2024publication} examines paper-volume growth and prolific-author patterns across the same venues. Our work adapts these methods to a regional venue and combines multiple data sources within a single corpus, enabling cross-cutting analyses that are typically studied in isolation.

\paragraph{Reproducibility, openness, and citation impact.}
A second line of work investigates how open research practices relate to citation impact. \citet{zhou2023impact} document that code-release rates roughly doubled over six years in machine learning, robotics, and control venues, providing the closest direct analog to our artifact analysis. \citet{goyal2023papers} estimate a citation-rate advantage of approximately 20\% following the creation of a GitHub repository, using nearest-neighbour matching to control for confounders. \citet{pineau2021reproducibility} offers the canonical account of community-level reproducibility initiatives at NeurIPS, and \citet{moed2007preprint} provides early evidence of the citation advantage associated with arXiv preprints. In contrast with this literature, we find that at BRACIS the rising openness signals are not accompanied by a robust citation premium once outlier Portuguese-NLP papers are removed, echoing concerns that observed premiums may partly reflect selection and promotion effects rather than openness itself \citep{arxivpromotion2024,kapoor2023leakage}.

\paragraph{Meta-science of Brazilian conferences.}
The Brazilian computing community has a long tradition of studying its own venues, with bibliometric analyses of CSBC \citep{limafilho2023csbcset}, SBSI \citep{carvalho2024sbsi}, WCAMA \citep{carvalho2024wcama}, SBBD \citep{procopio2017sbbd}, SBES \citep{mendonca2022sbes}, and SBRC \citep{vazdemelo2013sbrc}. \citet{dalpizzol2022gender} perform a name-based gender analysis across 13 SBC conferences, including two BRACIS editions, and \citet{albuquerque2024nlp} map NLP papers across BRACIS, BraSNAM, ENIAC, PROPOR, and STIL. The closest work to ours is the longitudinal study of the PROPOR ecosystem by \citet{nunes2026propor}, which analyzes the biennial conference from 2003 to 2024 and reports the same heavy-tailed citation pattern we observe at BRACIS. Our work complements this tradition along several dimensions that previous metadata-only studies could not address: public-artifact release, industry collaboration, LLM-topic adoption, reflective-section detection, and an intra-venue citation network, all made possible by an LLM-assisted full-text pipeline.

\paragraph{Heavy-tailed citation distributions.}
Heavy-tailed citation distributions have been studied since \citet{price1976general}'s formulation of preferential attachment, with \citet{clauset2009powerlaw} establishing the standard methodology for power-law fitting. \citet{golosovsky2012runaway} argue that runaway events dominate the heavy tail of citation distributions \citep{brzezinski2015power}, a framing that matches our finding that one or two outlier Portuguese-NLP papers drive BRACIS's right-tail behaviour. Regarding data sources, \citet{martinmartin2018google} document the coverage tradeoffs of Google Scholar against Web of Science and Scopus, which is directly relevant to our cross-source comparison with the OpenAlex-based counts used by \citet{nunes2026propor}, and DBLP \citep{ley2002dblp} remains the canonical metadata reference for computer science venues.

\section{Data and Methods}\label{sec:method}

We built a per-paper record of every BRACIS paper from 2015 to 2025. The pipeline pulls metadata from DBLP, citation counts from Google Scholar through Serper, arXiv links from Semantic Scholar, and structured fields from the PDFs themselves. The rest of this section walks through each step. Verbatim LLM prompts are in the appendix~\ref{app:prompts}.

\paragraph{Compiling the list of accepted papers.}
DBLP \citep{ley2002dblp} is our source for the paper list. We parsed the per-volume XML at \texttt{dblp.org/db/conf/bracis/<volume>.xml} and pulled out title, authors, DOI, page range, and topical session heading. We ended up with 1,066 papers.

\paragraph{Recovering the PDFs.}
BRACIS proceedings are not open access. We downloaded each paper by hand: from IEEE Xplore for the 2015 to 2019 editions and Springer for the 2020 to 2025 LNAI volumes, both through institutional access available to the authors of this work. We extracted plain text from the PDFs with \texttt{pdftotext}. Of the 1{,}066, 20 could not be retrieved and are excluded from every analysis that requires the PDF, leaving 1{,}046 eligible papers. An additional 14 papers were manually rescued but produced text-extraction output too broken to reliably scan for URLs, section headings, or contribution-type cues; they remain in citation, institution, gender, and other metadata analyses but sit outside the 1{,}032-paper subset used by the artifact-link, reflective-section, and contribution-type classifiers.

\paragraph{Citations, preprints, and best-paper nominees.}
We queried Google Scholar through the Serper.dev API for each paper's citation count, verifiing the Scholar authors and year against the BRACIS record to ensure we were considering the right paper. We obtain the arXiv link for each paper from Semantic Scholar's DOI crosswalk. The best-paper nominees were manually recovered through inspection of old bracis websites, publications, or old presentation schedules; we were able to retrieve nominees for 9 of 11 editions, totaling 48; we could not recover the 2017 or 2018 lists.

\paragraph{Extracting structured fields with sabiazinho-4.}

We use an LLM, sabiazinho-4~\cite{sabia4_report}, to structure and infer multiple aspects of the retrieved papers, namely:
\begin{enumerate}
\item \textbf{Author affiliations.} The model extracts institutions named as affiliations of any author of the paper. Institutions are canonicalized and deduplicated, and finally, we extract the location of each institution using sabiazinho-4 with a web search tool.
\item \textbf{First-author gender}, Following other studies~\cite{lariviere2013bibliometrics, nunes2026propor}, we infer the gender of the authors given their first name using sabiazinho-4.
\item \textbf{Topic and keywords.} Each paper is assigned one major-area label from a 14-class taxonomy covering Machine Learning, Natural Language Processing, Computer Vision, Optimization and Metaheuristics, Data Mining, Bioinformatics and Healthcare AI, Reinforcement Learning, Knowledge Representation and Reasoning, Multi-Agent Systems, Robotics, AI Ethics, Deep Learning and an \emph{unclassified} catch-all. In parallel, the model emits an open-vocabulary set of fine-grained keywords per paper (e.g.\ \texttt{portuguese-language}, \texttt{bert}, \texttt{transformers}) capturing the specific method, dataset, or domain focus. The keyword vocabulary is not constrained to a fixed list, so the LLM tends to emit lexical variants of the same concept; we apply a conservative canonicalization pass that merges singular/plural forms and a small hand-curated synonym table before counting, which leaves 2,646 keywords between all available papers. The top-30 canonical keywords are listed in Appendix~\ref{app:subtags}.
\item \textbf{Contribution-type and labels}, We also use the LLM to label the main contribution type of papers using a 4-class taxonomy.
\end{enumerate}

\paragraph{Taxonomy of the contribution types}
We assign each paper to one of four mutually-exclusive contribution types inspired by \citet{pramanick2025nature}:
\begin{itemize}
  \item \emph{Survey}: papers whose primary contribution is a literature review or position statement, including systematic mappings and methodological surveys.
  \item \emph{Model}: papers that propose a specific machine learning or neural network model, including new architectures, training schemes, embeddings, and pretrained-model releases such as BERTimbau and Sab\'ia.
  \item \emph{Algorithm}: papers that propose a new algorithm or a theoretical analysis, including search and optimization methods, metaheuristics, classical learning algorithms, and theoretical bounds.
  \item \emph{Empirical}: papers whose contribution is an empirical study, a benchmark or dataset, or the application of existing methods to a specific domain. Applications and tools sit here because BRACIS applications are evaluated empirically in practice and rarely propose a new method of their own.
\end{itemize}

\paragraph{Artifact links and reflective sections.}
To measure how often authors release artifacts, we run a regex sweep over the extracted text to find links to public repositories on GitHub, Hugging Face, GitLab, Bitbucket, Zenodo, OSF, and Kaggle. Similarly, to track the adoption of reflective writing, we apply regexes that detect the presence of \emph{Future Work} and \emph{Limitations} section headings.

\paragraph{Intra-BRACIS citation graph.}
We build a citation graph over BRACIS papers alone, which lets us study how often BRACIS papers cite earlier BRACIS work and whether a few hub papers anchor the venue's research. Each paper is a node, and we add a directed edge from one paper to another whenever the first cites the second. To find these edges, we extract the references from each paper in the corpus with two methods. A regex sweep locates each paper's references section by its heading and slides a six-line window through it, flagging an edge when at least 80\% of the cited paper's distinctive title tokens and the first-author surname appear together. The LLM pass sends the same references section to \texttt{sabiazinho-4} and asks for a structured list of citations with title, authors, year, and DOI when present; we then match each extracted reference to the corpus by exact DOI, by title-token Jaccard with a one-year tolerance, or by first-author surname, year, and title overlap. The final graph is the union of the two pipelines: 324 edges in total, 261 found by both methods, 37 by regex only, and 26 by the LLM only.

\paragraph{Institution collaboration graph.}
We also build a collaboration graph that treats each institution that appears on at least one BRACIS paper as a node and draws an edge between two institutions whenever they co-author at least one paper, with the edge weight equal to the number of co-authored papers. Institution names come from the canonicalized affiliation pass described above, so the graph operates on the deduplicated institution identities rather than on raw affiliation strings. We use this graph to study how often certain institutions collaborate.

\section{Results}\label{sec:results}

\subsection{Thematic Landscape: What?}\label{sec:rq1}

\textit{Which topics and contribution types define BRACIS, and how has the topical composition shifted across the decade?}

\begin{figure}[t]
  \centering
  \includegraphics[width=\linewidth]{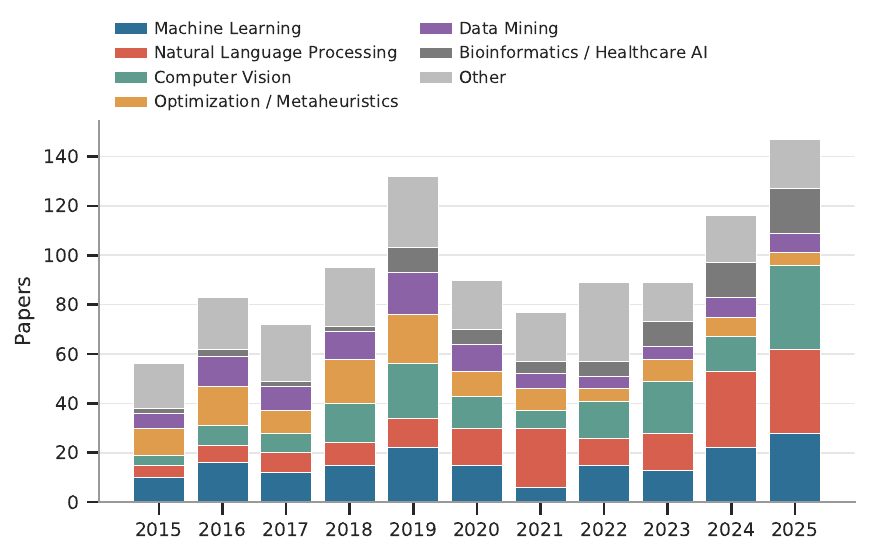}
  \caption{Per-year share of BRACIS papers by major area, 2015 to 2025.}
  \label{fig:topics}
\end{figure}

\textbf{Machine Learning, Natural Language Processing, and Computer Vision jointly account for about half of all BRACIS papers.}
Figure~\ref{fig:topics} shows the per-year paper counts for the six largest research areas across the 1,046 eligible papers, with all remaining areas grouped as \emph{Other}. Three areas lead the corpus at comparable scale: Machine Learning, Natural Language Processing, and Computer Vision, with 174, 171, and 162 papers respectively. These are followed by a substantial tier of Optimization and Metaheuristics, Data Mining, and Bioinformatics and Healthcare AI. Counting the next three areas not shown individually here, Reinforcement Learning, Knowledge Representation and Reasoning, and Multi-Agent Systems, these nine areas capture 87\% of all papers; the remaining 13\% spread across Robotics, AI Ethics, D, and Other Applications. The yearly breakdown reveals a diversified base that has not collapsed into any single area.

\begin{figure}[t]
  \centering
  \includegraphics[width=\linewidth]{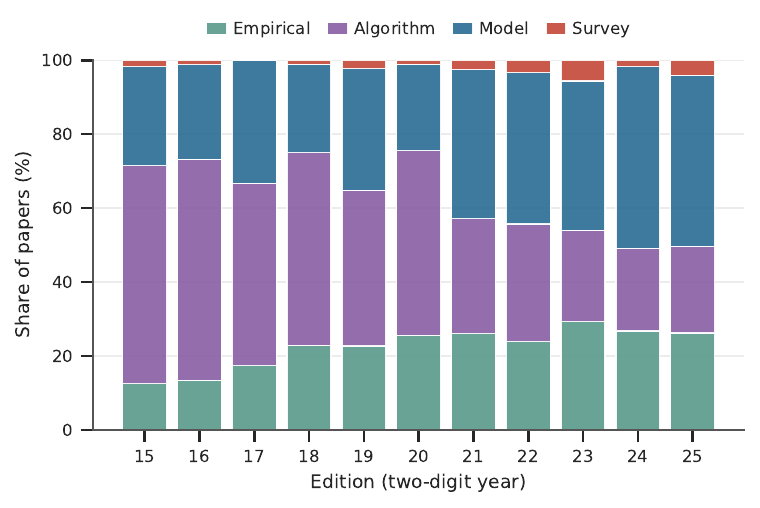}
  \caption{Per-year share of BRACIS papers by contribution group. Each paper is assigned its highest-priority group in the order Survey, Model, Algorithm, Empirical, so per-year shares sum to 100\%.}
  \label{fig:contribs}
\end{figure}

\textbf{Algorithm work has notably diminished, replaced by Model and Empirical contributions.}
Figure~\ref{fig:contribs} shows the per-year share of each contribution type, revealing a clear shift over the decade. Algorithm contributions fell from 59\% in 2015 to 23\% in 2025, while Model contributions rose from 27\% to 46\% and Empirical roughly doubled from 12\% to 26\%; Survey stayed small and roughly flat throughout. Across all 1,046 eligible papers, Algorithm remains the largest category overall at 39\%, followed by Model at 36\%, Empirical at 23\%, and Survey at 2\%. 

\begin{figure}[t]
  \centering
  \includegraphics[width=\linewidth]{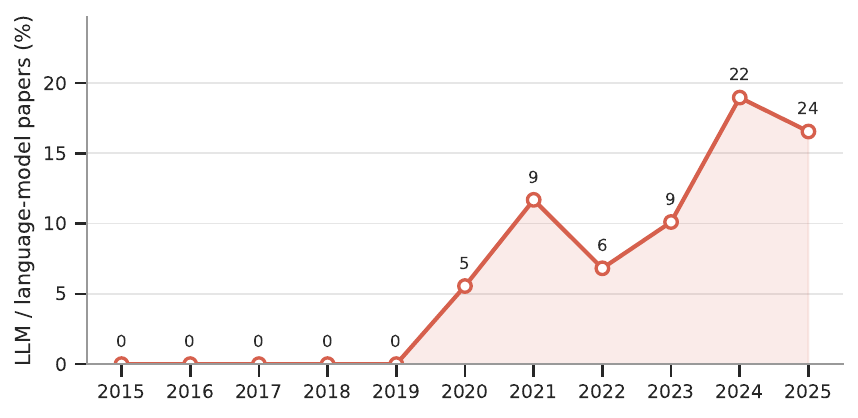}
  \caption{Per-year share of BRACIS papers tagged with LLM-related keywords such as \texttt{llm}, \texttt{bert}, \texttt{transformers}, and \texttt{large-language-models}. Numbers above markers are absolute paper counts.}
  \label{fig:llm-per-year}
\end{figure}

\textbf{Large Language Model research at BRACIS rose from zero before 2020 to 19\% by 2024.}
Figure~\ref{fig:llm-per-year} shows the trajectory. No paper before 2020 carries an LLM-related keyword such as \texttt{llm}, \texttt{bert}, or \texttt{large-language-models}. In 2020 the rate was 5.6\%; by 2024 it had reached 19\%.

\textbf{Portuguese-language and Brazilian-Portuguese keywords together rank among the top labels with 75 papers.}
Among the top fine-grained keywords, \texttt{portuguese-language} appears in 51 papers and \texttt{brazilian-portuguese} in 24, jointly placing language-specific work among the most common labels alongside \texttt{deep-learning} with 63 papers, \texttt{classification} with 52, The full top-30 keyword distribution is in Table~\ref{tab:subtags}. This concentration is consistent with the regional venue's role in absorbing Portuguese-language-specific research that larger international venues are unlikely to accept.

\subsection{The Community: Who?}\label{sec:rq2}

\textit{Who publishes at BRACIS, how do authors and institutions interact, and what does retention look like at each level?}

\begin{figure*}[t]
  \centering
  \includegraphics[width=0.95\linewidth]{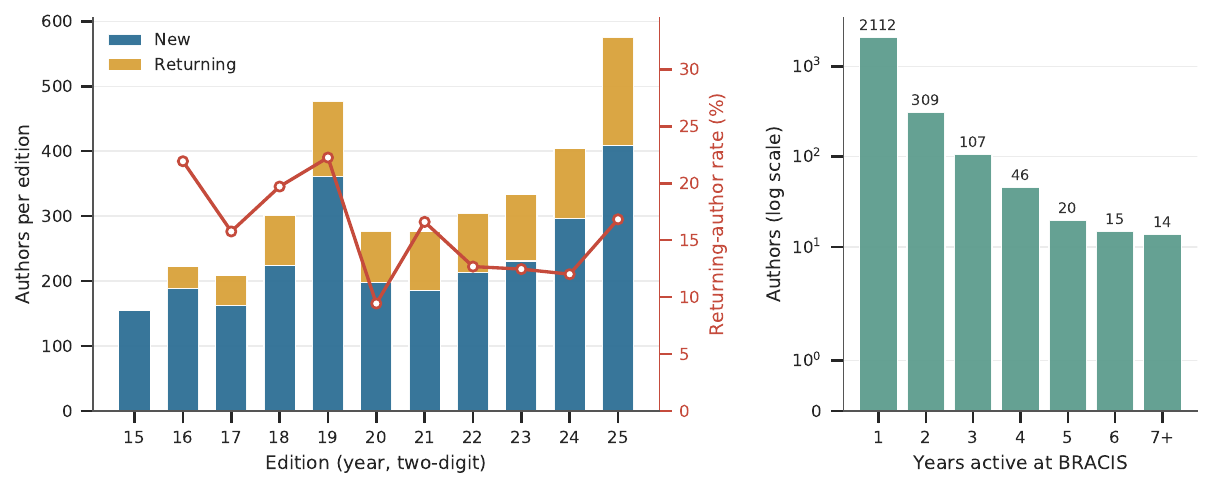}
  \caption{Author dynamics at BRACIS, 2015 to 2025. Left: new and returning authors stacked per edition with year-over-year retention rate on the right axis. Right: histogram of years-active per author, with the 7+ bucket aggregated.}
  \label{fig:author-dynamics}
\end{figure*}

\textbf{80.5\% of BRACIS authors appear in only one edition while year-over-year retention sits at 16\%.}
Figure~\ref{fig:author-dynamics} shows new versus returning authors per edition and the histogram of years-active per author. Across 1{,}066 papers we count 2{,}623 unique authors. 2{,}112 of them, or 80.5\%, appear in only one BRACIS edition, and just 49, or 1.9\%, appear in five or more editions. Year-over-year author retention ranges from 9\% in 2020 to 22\% in 2019 with a decade mean of 16\%, mirroring the hourglass pattern \citet{nunes2026propor} report for PROPOR with 15 to 25\% retention.

\begin{figure}[t]
  \centering
  \includegraphics[width=\linewidth]{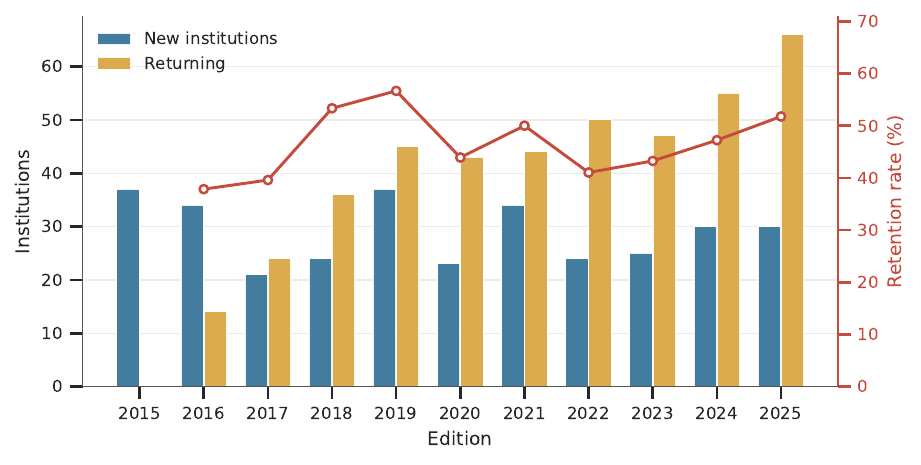}
  \caption{Institutional dynamics at BRACIS, 2015 to 2025. Bars show new versus returning institutions per edition; the red line shows year-over-year retention averaging 46.5\%.}
  \label{fig:inst-dynamics}
\end{figure}

\textbf{Institutional retention (46.5\%) is nearly three times author retention (16\%).}
Figure~\ref{fig:inst-dynamics} shows new versus returning institutions per edition with the retention curve. Per-edition institutional retention averages 46.5\% across 319 distinct institutions, nearly three times the 16\% author retention rate. This divergence corroborates the findings of \citet{nunes2026propor}: laboratories and universities provide structural continuity of the venue, while the individual contributor base turns over rapidly with each new student cohort.

\begin{figure}[t]
  \centering
  \includegraphics[width=\linewidth]{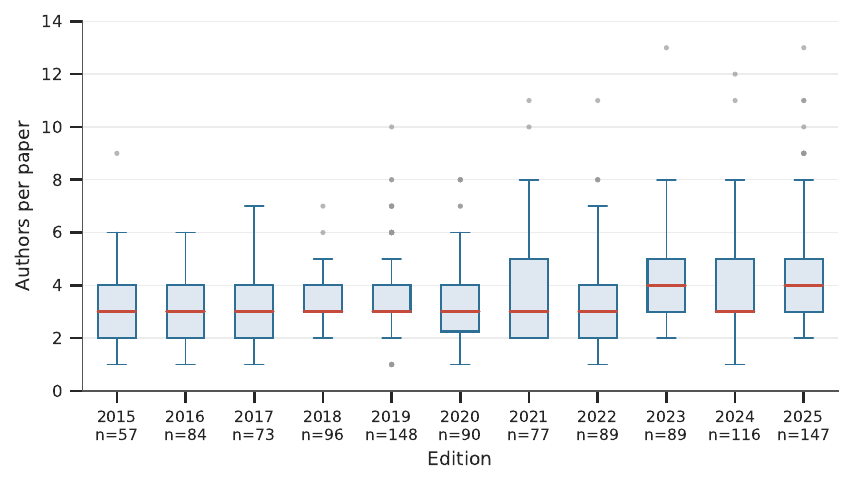}
  \caption{Per-year boxplot of authors per BRACIS paper. The median is 3 from 2015 to 2022 and 4 from 2023 onward.}
  \label{fig:team-size}
\end{figure}

\textbf{Team sizes have grown modestly across the decade.}
Figure~\ref{fig:team-size} shows the per-year distribution of authors per paper. The median paper had 3 authors from 2015 to 2022 and 4 from 2023 onward. The mean rose from 3.14 in 2015 to 4.41 in 2025, and the per-year maximum from 9 to 13. This is a milder version of the Big Science shift that \citet{nunes2026propor} observe at PROPOR; at BRACIS, the long upper tail of papers with at least 8 authors is concentrated in recent editions, but smaller teams remain the most common in the distribution.

\begin{figure}[t]
  \centering
  \includegraphics[width=\linewidth]{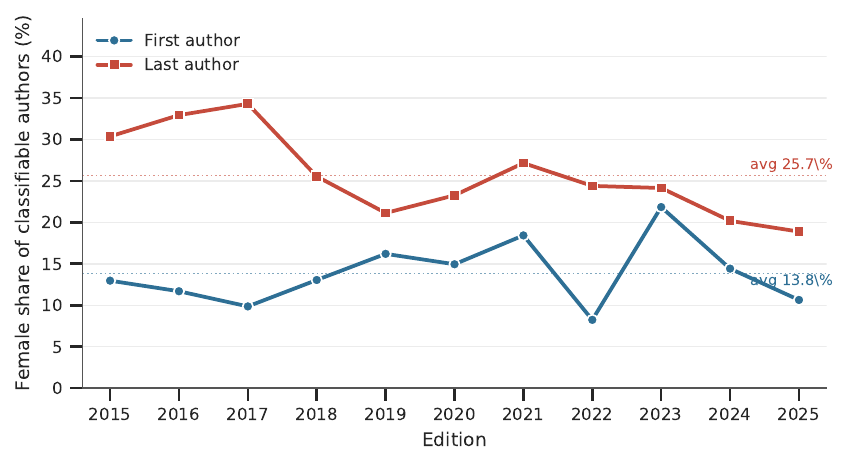}
  \caption{Female first-author share by BRACIS edition. The decade-long average is 13.3\% with no significant linear trend.}
  \label{fig:gender}
\end{figure}

\textbf{The rate of Female first authors is 14\%; while the female last author rate is consistently higher and declines over the decade.}
Figure~\ref{fig:gender} reports the per-edition distribution, across 1{,}066 papers; 13.3\% have a female first author and 23.6\% have a female last author. Among classifiable authors, the female first-author rate averages 13.8\% over the decade, ranging from 8.2\% in 2022 to 21.8\% in 2023, with no clear trend. The last-author position, conventionally the senior author and often the advisor in Brazilian doctoral and master's work, averages 25.7\% female and is higher than the first-author rate in every single edition. The last-author rate also drifts downward: it sits above 30\% in 2015 to 2017 and below 25\% in every edition since 2018, ending at 18.9\% in 2025. For context, \citet{mohammad2020state} reports a first-author rate near 29\% across the ACL Anthology, and \citet{dalpizzol2022gender} reports rates broadly in the same range we find at BRACIS for 13 Brazilian SBC conferences.

\begin{figure}[t]
  \centering
  \includegraphics[width=\linewidth]{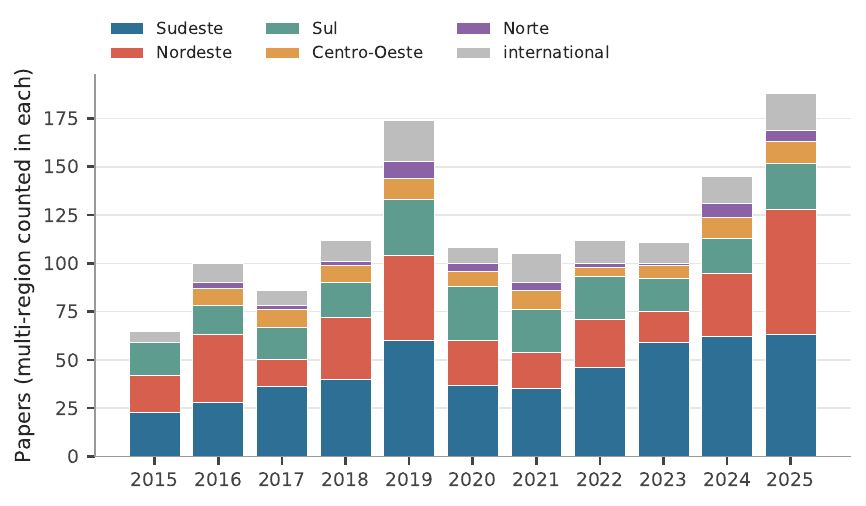}
  \caption{Per-year regional distribution of BRACIS papers. A paper is counted in each region from which it has at least one author affiliation.}
  \label{fig:regions}
\end{figure}

\textbf{The Brazilian Northeast region is the fastest-growing region at BRACIS.}
Overall, institutions from the Sudeste contribute in 46\% of BRACIS papers through the 10 years, Nordeste in 30\%, Sul 21\%, Centro-Oeste 8\%, and Norte 4\%, and with 13\% of papers featuring at least one international co-author. Figure~\ref{fig:regions} shows the per-year regional distribution. Nordeste participation grew 3.4 times, from 19 papers in 2015 to 65 in 2025, outpacing the overall conference growth of 2.6 times.

\begin{figure*}[t]
  \centering
  \includegraphics[width=\linewidth]{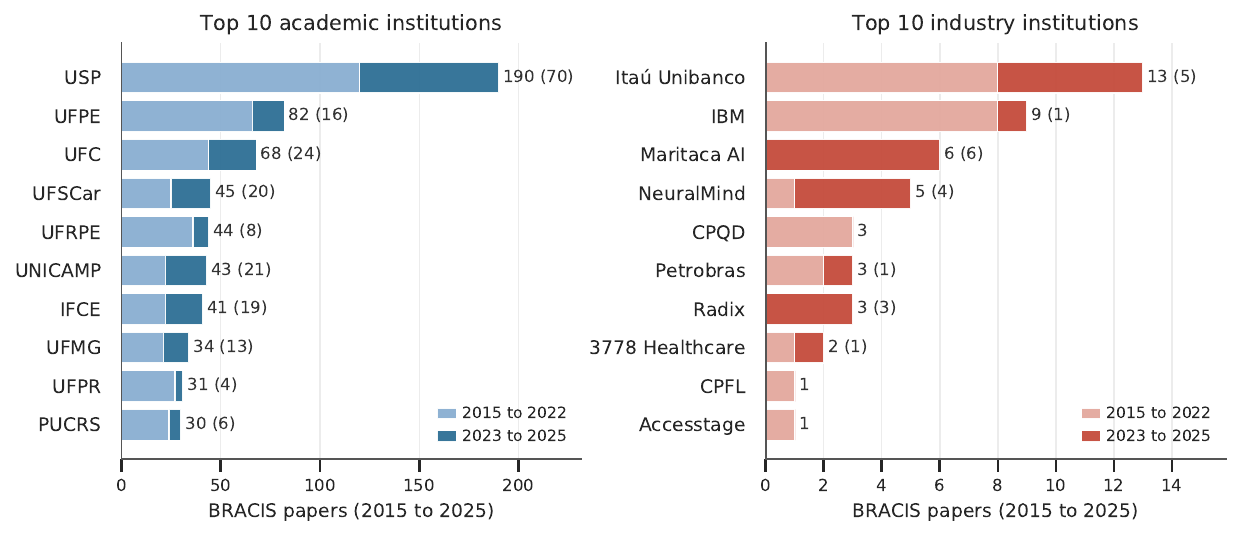}
  \caption{Top 10 academic and top 10 industry institutions by BRACIS paper count, 2015 to 2025. Each bar is split into an earlier-editions component (light, 2015 to 2022) and a recent-editions component (dark, 2023 to 2025); the annotation next to each bar is the total followed by the recent-editions count in parentheses. A paper is counted once per institution.}
  \label{fig:top-insts}
\end{figure*}

\textbf{USP leads volume with 190 papers; the top ten academic institutions are uniformly Brazilian public universities.}
Figure~\ref{fig:top-insts} compares the top 10 academic institutions and the top 10 industry institutions by BRACIS paper count. The academic ranking is led by USP with 190 papers, followed by UFPE with 82 and UFC with 68; all entries are Brazilian public universities. The industry ranking is dominated by Itaú Unibanco with 13 papers and IBM Research with 9, then the domestic Portuguese-NLP startups Maritaca AI with 6 and NeuralMind with 5.

\begin{figure}[t]
  \centering
  \includegraphics[width=0.95\linewidth]{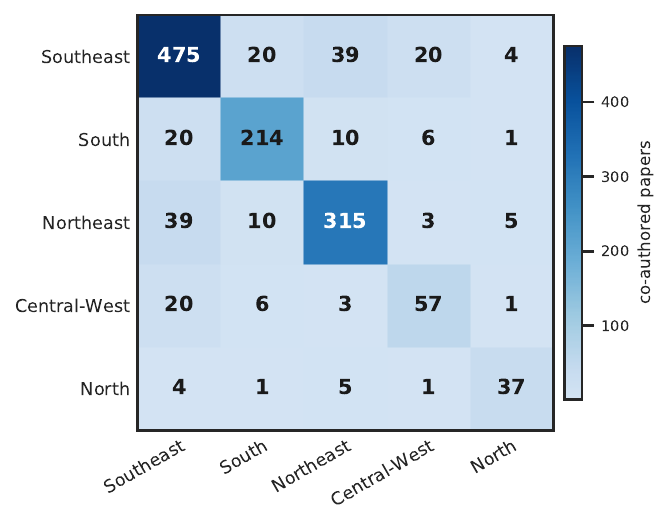}
  \caption{Region-by-region co-authorship matrix at BRACIS, 2015 to 2025. Cells count BRACIS papers whose authors span the row and column macroregions; diagonal counts are within-region collaborations.}
  \label{fig:region-collab}
\end{figure}

\textbf{Co-authorship concentrates within Brazilian macroregions.}
Figure~\ref{fig:region-collab} summarises co-authorship at the region level. The diagonal dwarfs every off-diagonal cell: Southeast carries 475 within-region co-authored papers, Northeast 315, South 214. The largest cross-region cell sits at 39 papers between Southeast and Northeast; cross-region collaboration is otherwise rare.

\begin{figure}[t]
  \centering
  \includegraphics[width=0.95\linewidth]{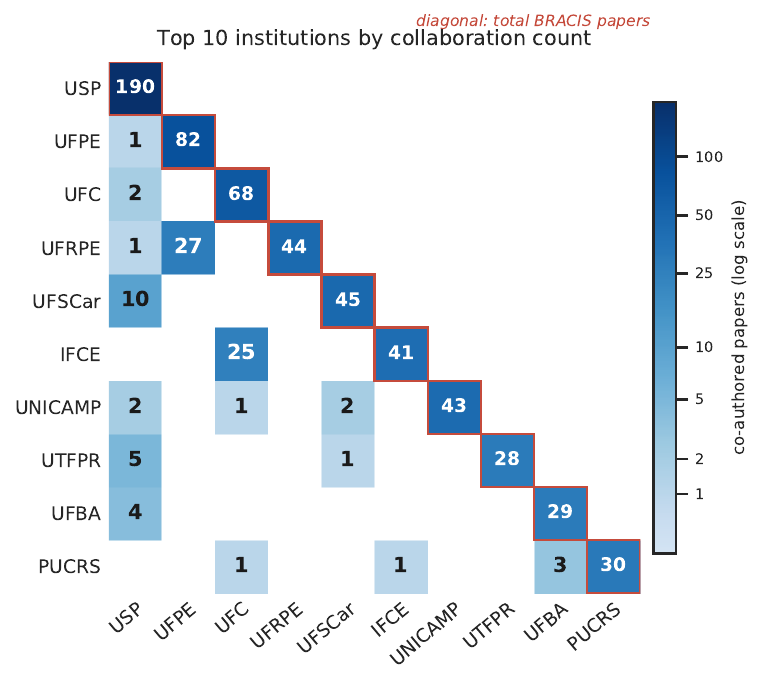}
  \caption{Co-authorship heatmap among the top 10 institutions by total collaboration count at BRACIS, 2015 to 2025. The matrix is symmetric and only the lower triangle is shown. Off-diagonal cells count co-authored BRACIS papers between the row and column institutions; diagonal cells, outlined in red, give each institution's total BRACIS paper count. Cell colour is on a log scale.}
  \label{fig:top-pairs}
\end{figure}

\textbf{Co-authorship among the top 10 institutions is sparse, with two intra-region anchor pairs dominating.}
Figure~\ref{fig:top-pairs} shows the 10$\times$10 co-authorship matrix for the top 10 institutions by total collaboration count. UFRPE-UFPE at 27 papers and IFCE-UFC at 25 are by far the strongest collaboration pairs and lie within the Northeast. USP-UFSCar at 10 is the only Southeast intra-region pair above 5. USP carries every other off-diagonal entry of 4 or more (USP-UTFPR 5, USP-UFBA 4), confirming its role as the venue's main cross-region connector. The matrix is otherwise sparse: most of the 90 off-diagonal cells are zero.

\begin{table}[t]
  \centering
  \small
  \caption{Top 10 institutions by h-index across the full BRACIS corpus, restricted to institutions with at least 5 papers. UNICAMP's $h=9$, $g=35$ gap is driven by BERTimbau.}
  \label{tab:hindex-full}
  \begin{tabular}{lrrrrr}
    \toprule
    Institution & $n$ & $h$ & $g$ & i10 & mean \\
    \midrule
    USP       & 190 & 16 & 27 & 32 &  6.3 \\
    UFPE      &  82 & 12 & 20 & 13 &  7.1 \\
    UNICAMP   &  43 &  9 & 35 &  8 & 28.9 \\
    PUCRS     &  30 &  9 & 15 &  7 &  8.3 \\
    UFSCar    &  45 &  8 & 17 &  6 &  7.0 \\
    UFC       &  68 &  8 & 16 &  7 &  4.4 \\
    UFRPE     &  44 &  8 & 13 &  7 &  6.4 \\
    UFPR      &  31 &  8 & 10 &  7 &  5.4 \\
    UFRN      &  27 &  8 & 10 &  5 &  4.7 \\
    UFU       &  28 &  7 & 12 &  5 &  6.5 \\
    \bottomrule
  \end{tabular}
\end{table}

\begin{table}[t]
  \centering
  \small
  \caption{Top 10 institutions by h-index over the last three BRACIS editions (2023 to 2025), restricted to institutions with at least 3 papers. Maritaca AI and NeuralMind enter the top-10 ranking.}
  \label{tab:hindex-recent}
  \begin{tabular}{lrrrr}
    \toprule
    Institution & $n$ & $h$ & $g$ & median \\
    \midrule
    UNICAMP     & 21 & 6 &  8 &  3 \\
    USP         & 70 & 5 &  7 &  0 \\
    Maritaca AI &  6 & 5 &  6 & 15 \\
    UFMG        & 13 & 4 &  4 &  1 \\
    UFSC        & 16 & 3 &  6 &  2 \\
    NeuralMind  &  4 & 3 &  4 & 15 \\
    UFOP        & 10 & 3 &  4 &  1 \\
    UFSCar      & 20 & 3 &  3 &  0 \\
    UFU         &  9 & 3 &  3 &  2 \\
    UFRGS       &  9 & 3 &  3 &  1 \\
    \bottomrule
  \end{tabular}
\end{table}

\subsection{Scientific Impact: What is the Impact?}\label{sec:rq3}

\textit{How are citations distributed across BRACIS papers, what predicts impact, and how do openness signals correlate with it?}

\begin{figure}[t]
  \centering
  \includegraphics[width=\linewidth]{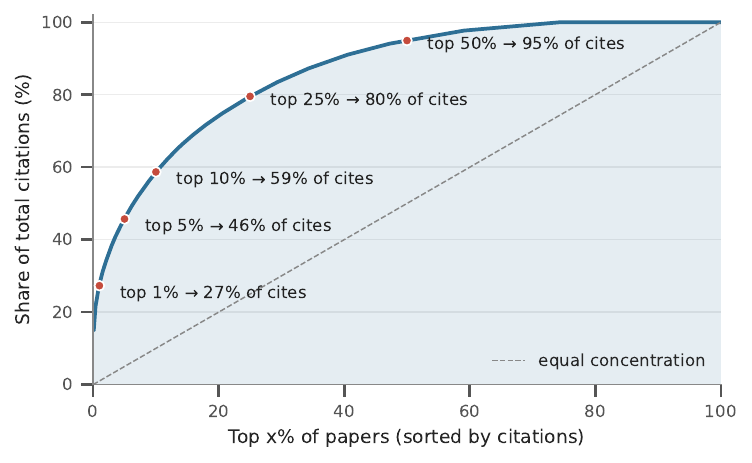}
  \caption{Citation concentration across the 1{,}046 eligible BRACIS papers. The top 1\% of papers carry 27\% of all citations and the top 10\% carry 58\%.}
  \label{fig:cite-dist}
\end{figure}

\begin{figure}[t]
  \centering
  \includegraphics[width=\linewidth]{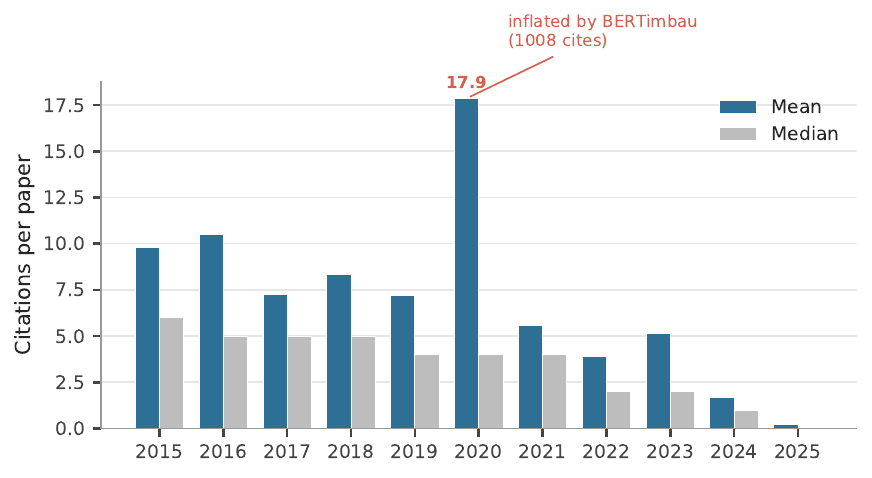}
  \caption{Per-year mean and median citation count for BRACIS papers. The decade-long mean of 6.5 is dominated by a small number of right-tail papers.}
  \label{fig:cites-per-year}
\end{figure}

\textbf{BRACIS citations are heavily skewed: the top 1\% of papers carry 27\% of all citations.}
Across the corpus, we count 6{,}765 total citations on 1{,}046 eligible papers. The mean is 6.5, and the median is 2, with 26\% of the analyzed papers receiving no citations; considering only papers at least three years old, 12\% remain uncited, indicating that low rates are not purely an age artifact. Figure~\ref{fig:cite-dist} plots the citation concentration curve and Figure~\ref{fig:cites-per-year} the per-year mean and median; Table~\ref{tab:peryear} gives the full per-year statistics. The top 1\% of papers carry 27\% of all citations and the top 10\% carry 59\%. The mean sits between the 75th and 80th percentiles.

\begin{table*}[!t]
  \centering
  \small
  \caption{Citation impact by openness signal for BRACIS papers at least three years old. ``+'' marks papers with the signal present; ``$-$'' marks those without. MW $p$ is a two-sided Mann-Whitney test; MW $p^\dagger$ is the same test after dropping all papers with at least 100 citations from both groups.}
  \label{tab:openness}
  \begin{tabular}{lrrrrrrrr}
    \toprule
    Signal      & $n_+$ & Mean$_+$ & Med$_+$ & $n_-$ & Mean$_-$ & Med$_-$ & MW $p$ & MW $p^\dagger$ \\
    \midrule
    arXiv       &  44 & 13.2 & 10 & 650 & 8.5 & 4 & $5{\times}10^{-5}$ & $4{\times}10^{-5}$ \\
    Artifact    & 208 & 12.5 &  4 & 486 & 7.2 & 4 & 0.47 & 0.47 \\
    Industry    &  40 & 30.0 &  4 & 654 & 7.5 & 4 & 0.78 & 0.60 \\
    \bottomrule
  \end{tabular}
\end{table*}

\begin{table*}[!t]
  \centering
  \small
  \caption{Citation comparison between best-paper nominees and non-nominees, restricted to the 9 editions with nominee data (2015, 2016, 2019, 2020, 2021, 2023, 2024, 2025; the 2022 nominees were recovered manually). Mann-Whitney $p$ tests the null hypothesis that the two distributions are equal.}
  \label{tab:nominees}
  \begin{tabular}{lrrrrr}
    \toprule
    Group & $n$ & Mean & Median & Max & MW $p$ \\
    \midrule
    Nominees                  &  48 & 26.0 & 3 & 1008 & 0.08 \\
    Nominees, excl.\ BERTimbau &  47 &  5.1 & 3 &   33 & 0.13 \\
    Non-nominees              & 849 &  5.1 & 2 &  166 & n/a \\
    \bottomrule
  \end{tabular}
\end{table*}

\begin{table}[!t]
  \centering
  \small
  \caption{Citation impact by contribution group for BRACIS papers at least three years old (year $\leq$ 2022, $n=682$: the 694 eligible papers from 2015 to 2022 minus 12 with no valid contribution-type label). MW $p$ is a two-sided Mann-Whitney test against the Algorithm group, the largest classical baseline. Algorithm itself shows the in-group statistics with no test.}
  \label{tab:contribcites}
  \begin{tabular}{lrrrrrr}
    \toprule
    Group     & $n$ & Mean & Med. &  Max & $h$ & MW $p$ \\
    \midrule
    Algorithm & 315 &  6.3 & 4 &  166 & 17 & n/a   \\
    Model     & 211 & 11.7 & 4 & 1008 & 18 & 0.82  \\
    Empirical & 144 &  9.6 & 6 &  134 & 20 & 0.004 \\
    Survey    &  12 & 10.5 & 6 &   32 &  6 & 0.064 \\
    \bottomrule
  \end{tabular}
\end{table}

\textbf{Treated as a single entity, BRACIS has h-index 30, g-index 56, and i10 162.}
Across 1{,}046 eligible papers, the corpus h-index sits at 30, meaning 30 papers have at least 30 citations each, the g-index at 56 with $g \approx 1.9\times h$, and 162 papers have at least 10 citations. Only four papers exceed 100 citations: BERTimbau \citep{souza2020bertimbau} with 1{,}008, Hyper-Parameter Tuning of Decision Trees with 166, Sab\'ia \citep{pires2023sabia} with 153, and Hate Speech Classification with 134. 

\textbf{The most-cited papers cluster on Portuguese-language NLP and on older applied deep learning.}
Table~\ref{tab:top20} lists the top 20 most-cited BRACIS papers. The list mixes two main patterns: Portuguese-language NLP releases including BERTimbau, Sab\'ia, Hate Speech in Social Media, Portuguese NER, and Topic Models for Brazilian Politics; and older applied deep-learning work such as offline signature verification, hard-drive failure prediction, and medicinal plant classification. The pre-2020 era is overrepresented: 17 of the top 20 papers are from 2015 to 2020, reflecting a citation lag, as more recent papers have not yet had time to accumulate citations.

\begin{table*}[t]
  \centering
  \small
  \caption{Top 20 most-cited BRACIS papers, according to Google Scholar counts as of May 2026. Some paper titles might be truncated.}
  \label{tab:top20}
  \begin{tabular}{rlrl}
    \toprule
    Rank & Year & Cit. & Title \\
    \midrule
     1 & 2020 & 1008 & \citet{souza2020bertimbau}--BERTimbau: Pretrained BERT Models for Brazilian Portuguese \\
     2 & 2016 &  166 & \citet{mantovani2016hyper}--Hyper-Parameter Tuning of a Decision Tree Induction Algorithm \\
     3 & 2023 &  153 & \citet{pires2023sabia}--Sab\'ia: Portuguese Large Language Models \\
     4 & 2018 &  134 & \citet{martins2018hate}--Hate Speech Classification in Social Media Using... \\
     5 & 2018 &   71 & \citet{souza2018writer}--A Writer-Independent Approach for Offline Signature Verification... \\
     6 & 2019 &   67 & \citet{da2019survey}--A Survey and Comparison of Trajectory Classification Methods \\
     7 & 2015 &   63 & \citet{cruz2015grouping}--Grouping Similar Trajectories for Carpooling Purposes \\
     8 & 2016 &   62 & \citet{glatt2016towards}--Towards Knowledge Transfer in Deep Reinforcement Learning \\
     9 & 2017 &   62 & \citet{dos2017predicting}--Predicting Failures in Hard Drives with LSTM Networks \\
    10 & 2019 &   61 & \citet{faustini2019fake}--Fake News Detection Using One-Class Classification \\
    11 & 2022 &   48 & \citet{fontanari2022cross}--Cross-validation Strategies for Balanced and Imbalanced Datasets \\
    12 & 2016 &   47 & \citet{marinho2016authorship}--Authorship Attribution via Network Motifs Identification \\
    13 & 2016 &   47 & \citet{chaves2016banhfap}--BaNHFaP: A Bayesian Network Based Failure Prediction Approach... \\
    14 & 2019 &   47 & \citet{pacifico2019automatic}--Automatic Classification of Medicinal Plant Species Based... \\
    15 & 2019 &   46 & \citet{santos2019assessing}--Assessing the Impact of Contextual Embeddings for Portuguese... \\
    16 & 2020 &   40 & \citet{nunes2020neural}Neural Architecture Search in Graph Neural Networks \\
    17 & 2021 &   38 & \citet{silva2021evaluating}Evaluating Topic Models in Portuguese Political Comments \\
    18 & 2020 &   37 & \citet{lochter2020deep}--Deep Learning Models for Representing Out-of-Vocabulary Words \\
    19 & 2020 &   37 & \citet{arruda2020measuring}--Measuring Instance Hardness Using Data Complexity Measures \\
    20 & 2020 &   36 & \citet{queiroz2020pre}--Pre-trained Data Augmentation... \\
    \bottomrule
  \end{tabular}
\end{table*}

\begin{figure}[t]
  \centering
  \includegraphics[width=\linewidth]{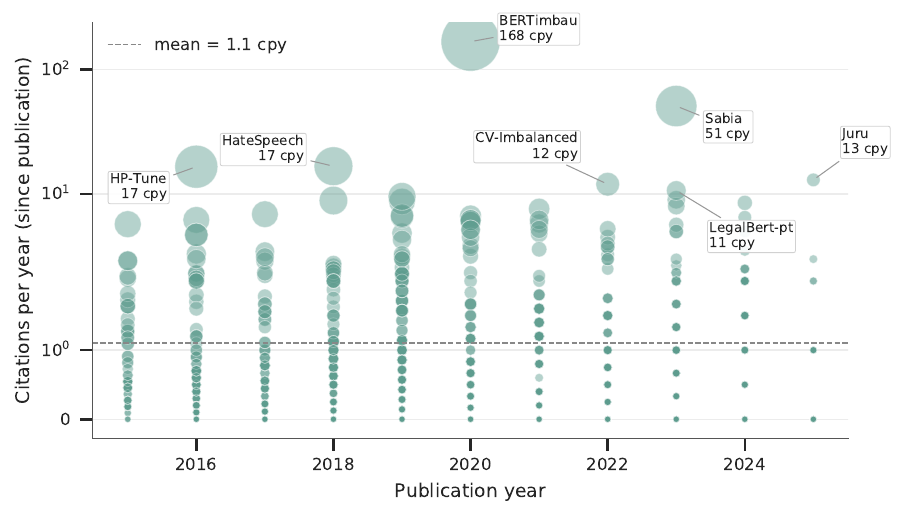}
  \caption{Citations-per-year (cpy) by publication year across the 1{,}046 eligible BRACIS papers, computed as total citations divided by integer years since publication. Marker size is proportional to total citation count. The dashed line is the corpus mean of 1.1 cpy.}
  \label{fig:cpy}
\end{figure}

\textbf{Citation velocity surfaces recent Portuguese-NLP releases that the top-20 table cannot yet see.}
Figure~\ref{fig:cpy} plots citations-per-year (cpy) by publication year. BERTimbau leads at 168 cpy and Sab\'ia at 51, both well above the corpus mean of 1.1. The Hate Speech and Hyper-Parameter Tuning papers reach 16.8 and 16.6 cpy through sustained accumulation over six to ten years. Three recent Portuguese-NLP releases that have not yet had time to enter the top-20 by total citations already exceed 8 cpy: Juru~\citet{junior2025juru} with 13.0 (a 2025 Brazilian legal LLM), LegalBert-pt \citet{silveira2023legalbert} with 10.7 (a 2023 legal language model), and ptt5-v2\citet{piau2024ptt5} with 8.5 (a 2024 continued-pretraining study). At those rates, several will likely enter the top 20 within two to three years.

\textbf{Intra-BRACIS citations are rare and concentrated on Portuguese-NLP hubs.}
We identified 324 verified intra-BRACIS edges from a hybrid regex + LLM pipeline (\S\ref{sec:method}). 22\% of papers cite at least one other BRACIS paper; the mean intra-venue in-degree is 0.31, the maximum in-degree is 41 (BERTimbau) and the maximum out-degree is 5, and the citation graph splits into 92 weakly connected components, with the largest reaching 79 papers. BERTimbau alone is cited by 41 other BRACIS papers, while the second-place Sab\'ia is cited by 10; outside these two hubs, the network carries only short chains. Table~\ref{tab:cite-network} reports the network-wide metrics and the most-cited papers within BRACIS.

\begin{table}[t]
  \centering
  \small
  \caption{Intra-BRACIS citation network. The top block summarises the directed graph of paper-to-paper citations; the bottom block lists the most-cited papers within the venue, with in-degree (citations received from other BRACIS papers) and out-degree (citations made to other BRACIS papers).}
  \label{tab:cite-network}
  \begin{tabular}{lr}
    \toprule
    \multicolumn{2}{l}{\emph{Network-wide}} \\
    \midrule
    Verified intra-BRACIS edges        & 324 \\
    Papers citing at least one (any)   & 233 \\
    Papers cited at least once (any)   & 187 \\
    Share of eligible papers citing    & 22\% \\
    Mean intra-venue in-degree         & 0.31 \\
    Max intra-venue in-degree          & 41 \\
    Max intra-venue out-degree         & 5 \\
    Weakly connected components        & 92 \\
    Largest component (papers)         & 79 \\
    \midrule
    \multicolumn{2}{l}{\emph{Most-cited papers within BRACIS (in / out)}} \\
    \midrule
    BERTimbau, 2020                    & 41 / 1 \\
    Sab\'ia, 2023                      & 10 / 4 \\
    Topic Models in Pt politics, 2021  & 5 / 1 \\
    Argumentation in MAS, 2015         & 4 / 0 \\
    Knowledge Rep.\ for Arg., 2016     & 4 / 1 \\
    NN Architectures in Quantum, 2017  & 4 / 0 \\
    Fault Detection in HDDs, 2016      & 4 / 1 \\
    Adaptive Operator NSGA-III, 2016   & 4 / 0 \\
    Portuguese NER Embeddings, 2019    & 4 / 0 \\
    Choquet Integral Apps, 2021        & 4 / 1 \\
    \bottomrule
  \end{tabular}
\end{table}

\begin{figure}[t]
  \centering
  \includegraphics[width=\linewidth]{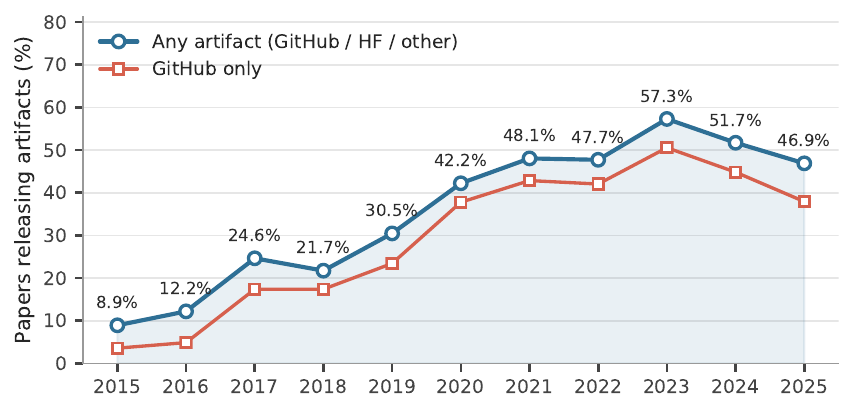}
  \caption{Per-year share of BRACIS papers linking to a public code or data artifact.}
  \label{fig:artifact}
\end{figure}

\textbf{Artifact sharing rose from 8.9\% in 2015 to 57.3\% in 2023.}
37.5\% of eligible papers, or 387 of 1{,}032, link to at least one public code or data artifact: 31\% to GitHub, 4.3\% to Hugging Face, and 10\% to Zenodo, OSF, Kaggle, or GitLab. Figure~\ref{fig:artifact} shows the per-year share rising from 8.9\% in 2015 to a peak of 57.3\% in 2023 (Spearman $\rho=0.88$, $p<10^{-3}$), broadly in line with the doubling trend that \citet{zhou2023impact} documents across international ML venues. Hugging Face URLs first appeared in 2020 alongside the LLM cohort.

\begin{figure}[t]
  \centering
  \includegraphics[width=\linewidth]{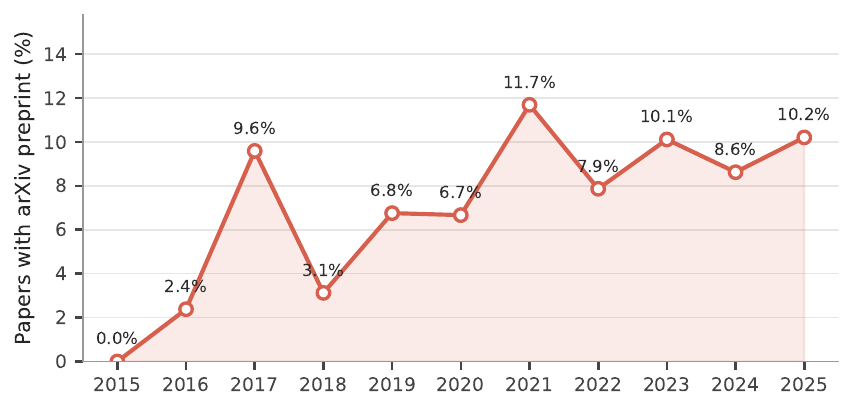}
  \caption{Per-year share of BRACIS papers with an arXiv preprint.}
  \label{fig:arxiv}
\end{figure}

\textbf{arXiv adoption at BRACIS reaches 7.4\%, , well below the rates at top international AI venues.}
Only 79 of 1{,}066 papers, or 7.4\%, have an arXiv preprint. Figure~\ref{fig:arxiv} shows the per-year rate at 0\% in 2015 and 10\% in 2025. arXiv adoption at BRACIS is therefore, far below the rates at international AI venues, at ICML 2023, over half of submissions appeared on arXiv before the review process concluded~\citet{su2025find}, and as early as 2017 the arXiv e-print rate reached 65\% at ICML and 57--64\% at other top ML and theory venues~\citet{sutton2017popularity}.

\begin{figure}[t]
  \centering
  \includegraphics[width=\linewidth]{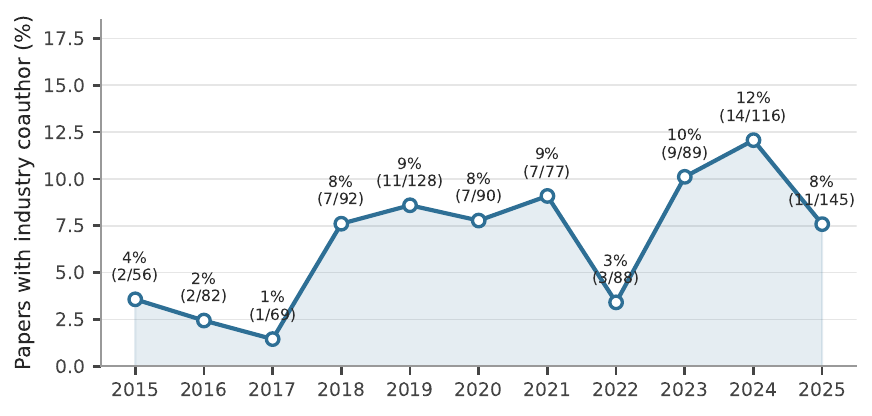}
  \caption{Per-year share of BRACIS papers with at least one industry-affiliated co-author.}
  \label{fig:industry}
\end{figure}

\textbf{Industry coauthorship has roughly doubled over the decade but remains rare.}
74 of 1{,}032 papers, or 7.2\%, list at least one industry-affiliated co-author. Figure~\ref{fig:industry} shows the per-year share rising from 3.6\% in 2015 to 7.6\% in 2025. The top industry contributors by paper count are the same set reported in Section~\ref{sec:rq2}: Ita\'u Unibanco leads with 13 papers, followed by IBM Research (9), Maritaca AI (6), and NeuralMind (5); Ita\'u and Petrobras (3) reflect industrial-research mandates at large Brazilian firms, while Maritaca AI and NeuralMind are research-active Portuguese-NLP startups concentrated in the last three editions. Maritaca AI's entire six-paper BRACIS footprint sits in 2023 to 2025, and four of NeuralMind's five papers are also in that window, with the remaining one being BERTimbau in 2020.

\textbf{Adoption of reflective sections has more than tripled across the decade.}
31.7\% of eligible papers carry an explicit \emph{Future Work}, \emph{Limitations}, or \emph{Conclusion and Future Work} section heading. The per-year rate rose from 12.5\% in 2015 to 51.7\% in 2025, and the \emph{Limitations}-specific rate from 3.6\% to 18.6\%. Unlike artifact release, reflective-section adoption is not driven by a few outlier papers; the trend is broadly distributed across topics.

\textbf{Among openness signals, only arXiv preprint posting correlates with citations.}

Table~\ref{tab:openness} compares arXiv posting, artifact release, and industry coauthorship on papers at least three years old. arXiv-posted papers carry mean 13.2 citations against 8.5 for non-arXiv, with median 10 against 4 and Mann-Whitney $p<10^{-4}$. Artifact release and industry coauthorship show no such correlation. Their raw mean ratios of 1.7 and 4.0 look favourable, but both fail the Mann-Whitney test  ($p=0.47$ for artifact, $p=0.78$ for industry) and both have median 4, identical to the papers without the signal.

\textbf{Empirical papers have the highest median citation count;}
Table~\ref{tab:contribcites} compares citation impact across the four contribution groups for papers at least three years old, controlling for citation lag. Empirical papers have median 6 citations, well above the median 4 of Model and Algorithm; the comparison against Algorithm is significant at $p=0.004$ under a two-sided Mann-Whitney test. Model papers have the highest absolute mean (11.7) but only because of BERTimbau; removing it would reduce the Model group mean to 7.0, statistically indistinguishable from Algorithm. Survey papers lean high (median 6, mean 10.5) but the sample is too small ($n=12$) to reach significance ($p=0.064$). BRACIS's robust citation impact sits in empirical, benchmark, and application work, while Model-style contributions deliver one blockbuster outlier and otherwise track the corpus baseline.

\textbf{Best-paper nominee status is uncorrelated with eventual citation impact.}
We tagged 48 best-paper nominees across 9 of 11 editions. Table~\ref{tab:nominees} reports the comparison. Nominees have mean 26.0 citations against 5.1 for non-nominees, a 5.1 times ratio, but removing BERTimbau, a 2020 nominee, drops the nominee mean to 5.1, identical to non-nominees. A Mann-Whitney test fails to reject equality of distributions even before outlier removal ($p=0.08$), and clearly so after ($p=0.13$). Medians sit at 3 for nominees and 2 for non-nominees in either case. Across all 9 years with nominee data, nominees appeared in only 2 of 45 possible top-5-most-cited slots. Female representation among nominees is 8.3\%, below the corpus-wide rate of 13.3\%, with six of nine years showing zero female-first-author nominees.

\section{Conclusion}\label{sec:conclusion}

Across eleven years and 1,066 papers, BRACIS emerges as an established regional venue with a distinct role in the Brazilian AI ecosystem. Our per-paper record lets us characterize the venue along three dimensions that prior metadata-only studies could not reach: what is published, who publishes it, and what impact the work accrues.

On the thematic side, BRACIS rests on a diversified base of Machine Learning, Natural Language Processing, and Computer Vision, with no single area dominating, while the contribution mix has shifted from Algorithm work toward Model and Empirical papers and LLM-related research has grown from zero before 2020 to roughly a fifth of recent editions. The community shows an hourglass structure: 80.5\% of authors appear in a single edition, but institutions provide continuity at nearly three times the author retention rate, with a stable core of Brazilian public universities anchoring the venue and the Northeast growing fastest among regions. Throughout, Portuguese-language work occupies a central place, consistent with BRACIS absorbing language-specific research that international venues are unlikely to publish.

Citation impact is heavily concentrated: the top 1\% of papers carry 27\% of all citations, and the right tail is dominated by a handful of Portuguese-language NLP releases. Openness practices have grown over the decade, but among them only arXiv preprint posting has a positive correlation with citations that survives outlier removal; artifact release and industry collaboration do not.

This points to access as a likely bottleneck. BRACIS proceedings sit behind IEEE and Springer paywalls, so a reader without a subscription reaches a paper through its preprint or not at all, yet only 7.4\% of papers have one. Closed proceedings paired with low arXiv adoption mean most BRACIS work is hard to reach from outside a subscribing institution, and the papers that do circulate are disproportionately the ones authors mirrored on arXiv. Moving future editions to an open-access model, as several SBC venues already publish through open proceedings, would remove this barrier and widen the reach of work published at the venue. 

Finally, we release the dataset and pipeline so the same analysis can be run on future editions and at other Brazilian AI conferences.

\section*{Limitations}\label{sec:limitations}

\paragraph{Citation source.}
Google Scholar's coverage is broader than alternatives \citep{martinmartin2018google}, but its API is unofficial, and indexing is opaque. Counts shift over time and across queries. Our verification pass corrects the most egregious false matches but cannot guarantee an unbiased census; absolute totals are conditional on a single May 2026 snapshot.

\paragraph{PDF and text extraction.}
34 papers are excluded from analyses that depend on full-text extraction (artifact URLs, reflective-section flags, contribution-type labels): 20 could not be retrieved at all and are dropped from every analysis, while another 14 were manually rescued but produced text-extraction output too broken to reliably scan and are excluded only from the full-text pipeline. This leaves 1{,}046 papers for citation, institution, and gender analyses and 1{,}032 papers for the full-text subset. \texttt{pdftotext}-based extraction is brittle on multi-column layouts and produces line-wrapped URLs that require pre-processing.

\paragraph{LLM classification errors.}
Institution canonicalisation occasionally mis-merged distinct institutions and was corrected with a manual fix-up dictionary. The contribution-type classifier was verified against the top-5 papers in each category but not audited at scale. We use \texttt{sabiazinho-4}, whose behaviour may drift; replication requires the same model or careful re-validation.

\paragraph{Gender inference.}
LLM-based name-to-gender classification is imperfect \citep{santamaria2018gender}. The 4\% unclassifiable rate captures initials and many international names; mid-confidence misclassification on Portuguese names cannot be excluded.

\paragraph{Causality.}
Every association we report between an openness signal and citation impact is correlational. Artifact release and industry collaboration premiums vanish after outlier removal, which already warns against causal interpretation; the arXiv preprint premium survives outlier removal but could still reflect selection effects, with authors who post on arXiv being a particular subset of the community.

\bibliographystyle{plainnat}
\bibliography{main}

\onecolumn
\appendix

\section{Top fine-grained keywords}\label{app:subtags}

Section~\ref{sec:rq1} reports the distribution of papers across the 14 top-level topic areas of our taxonomy. The classifier also emits free-form, kebab-case keywords on top of the area label, giving a much finer view of what each paper is actually about. Table~\ref{tab:subtags} lists the 30 most frequent keywords after singular/plural merging and the hand-curated synonym table described in Section~\ref{sec:rq1}. The list is dominated by methods rather than applications: \texttt{deep-learning}, \texttt{convolutional-neural-networks}, and \texttt{classification} sit at the top, while Portuguese-specific tags (\texttt{portuguese-language}, \texttt{brazilian-portuguese}) and LLM-era tags (\texttt{bert}, \texttt{large-language-models}, \texttt{transformers}) confirm the late-decade Portuguese-NLP cohort visible in Figure~\ref{fig:llm-per-year}.

\begin{table}[H]
  \centering
  \small
  \caption{Top 30 fine-grained keywords by frequency across 1{,}046 eligible papers, after canonicalisation of singular/plural variants and a hand-curated synonym table.}
  \label{tab:subtags}
  \begin{tabular}{lr@{\hspace{2em}}lr}
    \toprule
    Keyword & Count & Keyword & Count \\
    \midrule
    deep-learning                  & 63 & explainability               & 25 \\
    classification                 & 52 & multi-objective-optimization & 24 \\
    convolutional-neural-networks  & 52 & feature-selection            & 24 \\
    portuguese-language            & 51 & brazilian-portuguese         & 24 \\
    medical-imaging                & 46 & text-classification          & 23 \\
    machine-learning               & 43 & semi-supervised-learning     & 21 \\
    metaheuristics                 & 42 & transfer-learning            & 21 \\
    genetic-algorithm              & 39 & text-mining                  & 20 \\
    evolutionary-algorithms        & 37 & sentiment-analysis           & 20 \\
    time-series                    & 37 & fuzzy-logic                  & 20 \\
    clustering                     & 36 & named-entity-recognition     & 20 \\
    unsupervised-learning          & 34 & neural-networks              & 18 \\
    ensemble-learning              & 33 & graph-neural-networks        & 18 \\
    bert                           & 32 & meta-learning                & 17 \\
    large-language-models          & 32 & image-classification         & 31 \\
    \bottomrule
  \end{tabular}
\end{table}

\section{First-author gender by edition}\label{app:gender}

Figure~\ref{fig:gender} in Section~\ref{sec:rq2} summarises the female-share trajectory; Table~\ref{tab:gender} gives the raw counts behind it. We classify the first author of every paper as Male, Female, or Unknown using sabiazinho-4 on the given name. The Unknown bucket is small but non-zero in every edition: most cases are initials in place of given names, the rest are international names where Portuguese-first prompting cannot disambiguate. The F\% column normalises against classifiable authors only, so the rate is not depressed by Unknown counts.

\begin{table}[H]
  \centering
  \small
  \caption{First-author gender per BRACIS edition. \emph{F\%} is the female share of classifiable first authors. Row totals match the accepted-paper count for each edition.}
  \label{tab:gender}
  \begin{tabular}{lrrrrr}
    \toprule
    Year & Male & Female & Unkn. & Total & F\% \\
    \midrule
    2015 &  47 &  7 & 3 &   57 & 13.0 \\
    2016 &  68 &  9 & 7 &   84 & 11.7 \\
    2017 &  64 &  7 & 2 &   73 &  9.9 \\
    2018 &  80 & 12 & 4 &   96 & 13.0 \\
    2019 & 119 & 23 & 6 &  148 & 16.2 \\
    2020 &  74 & 13 & 3 &   90 & 14.9 \\
    2021 &  62 & 14 & 1 &   77 & 18.4 \\
    2022 &  78 &  7 & 4 &   89 &  8.2 \\
    2023 &  68 & 19 & 2 &   89 & 21.8 \\
    2024 &  95 & 16 & 5 &  116 & 14.4 \\
    2025 & 126 & 15 & 6 &  147 & 10.6 \\
    \midrule
    \textbf{All} & \textbf{881} & \textbf{142} & \textbf{43} & \textbf{1{,}066} & \textbf{13.9} \\
    \bottomrule
  \end{tabular}
\end{table}

\section{Per-year citation statistics}\label{app:peryear}

Section~\ref{sec:rq3} reports the corpus-level citation distribution. Table~\ref{tab:peryear} gives the per-edition breakdown so the headline mean of 6.5 can be read in context. Two patterns stand out. First, the 2020 edition's mean of 17.9 is more than twice any other year's, driven entirely by BERTimbau at 1{,}008 citations; the 2020 median sits at 4, identical to neighbouring years. Second, the right tail of the \emph{Zero} column grows mechanically with recency: 41 of 116 papers from 2024 and 128 of 147 papers from 2025 remain uncited at the time of the scrape, an artefact of the limited citation window rather than a quality signal. For papers at least three years old, the uncited share stabilises around 12\%.

\begin{table}[H]
  \centering
  \small
  \caption{Per-year volume and citation statistics across all 1{,}066 BRACIS papers, with verified Google Scholar counts scraped in May 2026. Citations come from Scholar via title and authors, so the PDF-eligibility filter applied elsewhere in the paper is not required here: the row-by-row means multiply out to 6{,}904 total citations across the 1{,}066 papers, of which the 1{,}046 eligible subset used in Section~\ref{sec:rq3} accounts for 6{,}765. \emph{Zero} counts papers with zero citations; $\geq\!10$ and $\geq\!50$ count papers above each threshold.}
  \label{tab:peryear}
  \begin{tabular}{rrrrrrrr}
    \toprule
    Year & $n$ & Mean & Med. & Max & Zero & $\geq 10$ & $\geq 50$ \\
    \midrule
    2015 &  57 &  9.8 & 6 &   63 &   3 & 21 & 1 \\
    2016 &  84 & 10.5 & 5 &  166 &   6 & 22 & 2 \\
    2017 &  73 &  7.3 & 5 &   62 &   7 & 17 & 1 \\
    2018 &  96 &  8.4 & 5 &  134 &   9 & 23 & 2 \\
    2019 & 148 &  7.2 & 4 &   67 &  15 & 35 & 2 \\
    2020 &  90 & 17.9 & 4 & 1{,}008 &  15 & 18 & 1 \\
    2021 &  77 &  5.6 & 4 &   38 &  10 &  8 & 0 \\
    2022 &  89 &  3.9 & 2 &   48 &  21 & 10 & 0 \\
    2023 &  89 &  5.1 & 2 &  153 &  20 &  7 & 1 \\
    2024 & 116 &  1.7 & 1 &   17 &  41 &  3 & 0 \\
    2025 & 147 &  0.2 & 0 &   13 & 128 &  1 & 0 \\
    \midrule
    \textbf{Total} & \textbf{1{,}066} & \textbf{6.5} & \textbf{2} & \textbf{1{,}008} & \textbf{275} & \textbf{165} & \textbf{10} \\
    \bottomrule
  \end{tabular}
\end{table}

\section{Most-cited paper per edition}\label{app:top-per-year}

Section~\ref{sec:rq3} reports the aggregate citation distribution and the venue-level top-20. Table~\ref{tab:top-per-year} cuts the same data by edition, showing the single most-cited paper from each year alongside the affiliations declared in the paper. The per-year peaks span an order of magnitude: BERTimbau (UNICAMP, NeuralMind, U.~Waterloo, 2020) leads the decade at over a thousand citations, the 2016 hyper-parameter tuning paper and the 2018 hate-speech study both cleared the hundred mark, and 2024 to 2025 entries are already accumulating despite their short citation window. Six of the eleven yearly peaks involve a São Paulo state institution (USP, UNICAMP, UFSCar, or Maritaca AI in Campinas), and from 2023 onward every peak features Maritaca AI as an author affiliation, mirroring the Portuguese-LLM concentration described in Section~\ref{sec:rq3}.

\begin{table}[H]
  \centering
  \small
  \caption{Most-cited BRACIS paper per edition, with citation count from Google Scholar (queried May 2026) and the affiliations declared in the paper. Affiliations use acronyms where available.}
  \label{tab:top-per-year}
  \begin{tabular}{l p{8.6cm} r p{4.6cm}}
    \toprule
    Year & Title & Cites & Affiliations \\
    \midrule
    2015 & \cite{cruz2015grouping}--Grouping Similar Trajectories for Carpooling Purposes & 63 & UFS \\
    2016 & \cite{mantovani2016hyper}--Hyper-Parameter Tuning of a Decision Tree Induction Algorithm & 166 & UFSCar, USP, U.~Pavla Jozefa \v{S}af\'arika, TU Eindhoven \\
    2017 & \cite{dos2017predicting}Predicting Failures in Hard Drives with LSTM Networks & 62 & UFC \\
    2018 & \cite{martins2018hate}--Hate Speech Classification in Social Media Using Emotional Analysis & 134 & U.~Minho \\
    2019 & \cite{da2019survey}--A Survey and Comparison of Trajectory Classification Methods & 67 & UFSC \\
    2020 & \cite{souza2020bertimbau}--BERTimbau: Pretrained BERT Models for Brazilian Portuguese & 1{,}008 & UNICAMP, NeuralMind, U.~Waterloo \\
    2021 & \cite{silva2021evaluating}--Evaluating Topic Models in Portuguese Political Comments About Bills from Brazil's Chamber of Deputies & 38 & UFG, UFU, USP, C\^amara dos Deputados \\
    2022 & \cite{fontanari2022cross}--Cross-validation Strategies for Balanced and Imbalanced Datasets & 48 & UFRGS, HCPA \\
    2023 & \cite{pires2023sabia}--Sabi\'a: Portuguese Large Language Models & 153 & Maritaca AI \\
    2024 & \cite{piau2024ptt5}--ptt5-v2: A Closer Look at Continued Pretraining of T5 Models for the Portuguese Language & 17 & UNICAMP, NeuralMind, Maritaca AI \\
    2025 & \cite{junior2025juru}--Juru: Legal Brazilian Large Language Model from Reputable Sources & 13 & USP, Maritaca AI \\
    \bottomrule
  \end{tabular}
\end{table}

\section{Recent leaderboard and citation velocity}\label{app:recent-cpy}

The venue-level top-20 in Section~\ref{sec:rq3} is dominated by pre-2020 papers because older work has had more time to accumulate citations. To surface the recent leaderboard and the fast-moving papers, we report two complementary cuts of the same citation data. Table~\ref{tab:recent-top10} lists the top 10 papers published in the last three editions (2023 to 2025) by total citations, using the same three-year window as the recent-impact analyses in Section~\ref{sec:rq2}; Table~\ref{tab:top-cpy} lists the top 10 papers of the whole decade by citations per year (CPY), which normalises for age.

The 2023-to-2025 ranking is heavily Portuguese-language NLP: Sab\'ia (Maritaca AI, 2023) leads at 153 citations, an order of magnitude ahead of the next entry, LegalBert-pt (IFCE, 2023) at 32. Portuguese-NLP resources (Sab\'ia, LegalBert-pt, BLUEX, ptt5-v2, InRanker, Juru) fill six of the ten slots. The CPY ranking is more revealing about future impact: BERTimbau (168 CPY) is followed by Sab\'ia at 51 CPY, still an order-of-magnitude gap; Juru (2025, 13.0 CPY) and LegalBert-pt (2023, 10.7 CPY) then confirm that the Portuguese-legal-LLM cluster is where recent citation momentum is concentrated.

\begin{table}[H]
  \centering
  \small
  \caption{Top 10 BRACIS papers published in the last three editions (2023 to 2025) by total Google Scholar citations, using the same window as the recent per-paper impact analyses in Section~\ref{sec:rq2}. Cites is the raw total; CPY is citations divided by integer years since publication (against a May 2026 snapshot).}
  \label{tab:recent-top10}
  \begin{tabular}{lp{6.2cm}rrp{3.1cm}}
    \toprule
    Year & Title & Cites & CPY & Affiliations \\
    \midrule
    2023 & \cite{pires2023sabia}--Sabi\'a: Portuguese Large Language Models & 153 & 51.0 & Maritaca AI \\
    2023 & \cite{silveira2023legalbert}--LegalBert-pt: A Pretrained Language Model for the Brazilian Legal Domain & 32 & 10.7 & IFCE \\
    2023 & \cite{rocha2023applying}--Applying Theory of Mind to Multi-agent Systems: A Systematic Review & 27 &  9.0 & KCL, UFSC \\
    2023 & \cite{pantoja2023spin}--A Spin-off Version of Jason for IoT and Embedded Multi-Agent Systems & 24 &  8.0 & CEFET/RJ, UFF \\
    2023 & \cite{almeida2023bluex}--BLUEX: A Benchmark Based on Brazilian Leading Universities Entrance Exams & 17 &  5.7 & UNICAMP, NeuralMind, Maritaca AI \\
    2024 & \cite{piau2024ptt5}--ptt5-v2: A Closer Look at Continued Pretraining of T5 Models for Portuguese & 17 &  8.5 & UNICAMP, NeuralMind, Maritaca AI \\
    2023 & \cite{sartori2023d}--d-CC Integrals: Generalizing CC-Integrals by Restricted Dissimilarity Functions & 15 &  5.0 & FURG, UFRN \\
    2023 & \cite{schiavon2023interpreting}--Interpreting Convolutional Neural Networks for Brain Tumor Classification & 15 &  5.0 & UFCSPA \\
    2024 & \cite{laitz2024inranker}--InRanker: Distilled Rankers for Zero-Shot Information Retrieval & 13 &  6.5 & UNICAMP, NeuralMind, Maritaca AI, Zeta Alpha \\
    2025 & \cite{junior2025juru}--Juru: Legal Brazilian Large Language Model from Reputable Sources & 13 & 13.0 & USP, Maritaca AI \\
    \bottomrule
  \end{tabular}
\end{table}

\begin{table}[H]
  \centering
  \small
  \caption{Top 10 BRACIS papers of the decade by citations per year (CPY), against the May 2026 Scholar snapshot. CPY = total citations divided by integer years since publication (matching the convention used in Section~\ref{sec:rq3} and Figure~\ref{fig:cpy}).}
  \label{tab:top-cpy}
  \begin{tabular}{lp{6.2cm}rrp{3.1cm}}
    \toprule
    Year & Title & Cites & CPY & Affiliations \\
    \midrule
    2020 & \cite{souza2020bertimbau}--BERTimbau: Pretrained BERT Models for Brazilian Portuguese & 1{,}008 & 168.0 & UNICAMP, NeuralMind, U.~Waterloo \\
    2023 & \cite{pires2023sabia}--Sabi\'a: Portuguese Large Language Models & 153 & 51.0 & Maritaca AI \\
    2018 & \cite{martins2018hate}--Hate Speech Classification in Social Media Using Emotional Analysis & 134 & 16.8 & U.~Minho \\
    2016 & \cite{mantovani2016hyper}--Hyper-Parameter Tuning of a Decision Tree Induction Algorithm & 166 & 16.6 & UFSCar, USP \\
    2025 & \cite{junior2025juru}--Juru: Legal Brazilian Large Language Model from Reputable Sources & 13 & 13.0 & USP, Maritaca AI \\
    2022 & \cite{fontanari2022cross}--Cross-validation Strategies for Balanced and Imbalanced Datasets & 48 & 12.0 & UFRGS \\
    2023 & \cite{silveira2023legalbert}--LegalBert-pt: A Pretrained Language Model for the Brazilian Legal Domain & 32 & 10.7 & IFCE \\
    2019 & \cite{da2019survey}--A Survey and Comparison of Trajectory Classification Methods & 67 & 9.6 & UFSC \\
    2023 & \cite{rocha2023applying}--Applying Theory of Mind to Multi-agent Systems: A Systematic Review & 27 & 9.0 & KCL, UFSC \\
    2018 & \cite{souza2018writer}--A Writer-Independent Approach for Offline Signature Verification & 71 & 8.9 & UFPE \\
    \bottomrule
  \end{tabular}
\end{table}

\section{Citation distribution percentiles}\label{app:percentiles}

Figure~\ref{fig:cite-dist} plots the citation concentration curve and Section~\ref{sec:rq3} reports the headline statistics (mean 6.5, median 2, top 1\% carrying 27\% of all citations). Table~\ref{tab:percentiles} gives the full numerical breakdown at standard percentiles so the curve can be read off directly. The numbers expose how skewed the distribution is: half of all BRACIS papers have 2 citations or fewer, three quarters have at most 6, and the jump from p95 to p99 (20 to 47) is larger than the jump from p50 to p95 (2 to 20). This is the same heavy-tail signature discussed in Section~\ref{sec:rq3} and matches the pattern \citet{nunes2026propor} report at PROPOR.

\begin{table}[H]
  \centering
  \small
  \caption{Citation count by percentile across the 1{,}046 eligible BRACIS papers (identical to the values on the full 1{,}066-paper corpus to the nearest integer). The mean of 6.5 sits between the 75th and 80th percentiles, a classic heavy-tail signature where a small number of high-impact papers pull the mean well above the typical paper.}
  \label{tab:percentiles}
  \begin{tabular}{lrrrrrrrrrrrr}
    \toprule
    Pct  & p10 & p20 & p30 & p40 & p50 & p60 & p70 & p75 & p80 & p90 & p95 & p99 \\
    \midrule
    Cites &  0 &  0 &  1 &  1 &  2 &  4 &  5 &  6 &  8 & 14 & 20 & 47 \\
    \bottomrule
  \end{tabular}
\end{table}

\section{Classification prompts}\label{app:prompts}

The topic-area and contribution-type labels reported in Section~\ref{sec:rq1} are produced by a single LLM (sabiazinho-4) at temperature 0, with one classifier per task. Both system prompts are shown below. The production prompts run against the corpus are written in Portuguese, since BRACIS papers are themselves largely in Portuguese and sabiazinho-4 is fine-tuned for Brazilian Portuguese; we reproduce here the literal English translation for readability, and ship the original Portuguese text alongside the inference code in the public repository.

\paragraph{Topic classification.} The classifier receives the paper title and the first 4{,}000 characters of the full text (typically the abstract plus introduction). It returns one of the 14 area labels listed in Section~\ref{sec:rq1} and a free-form list of 2 to 5 fine-grained keywords in lower-case-kebab-case. The system prompt is shown in Figure~\ref{fig:prompt-topic}.

\begin{figure}[H]
\centering
\fboxsep=0.7em
\fbox{\begin{minipage}{\dimexpr0.97\linewidth-2\fboxsep-2\fboxrule\relax}
\small\ttfamily\raggedright
You classify Artificial Intelligence research papers. Given the title and the beginning (abstract / introduction) of a paper, return EXACTLY one JSON with two keys:\par
\smallskip
- area: string, must be ONE of the options from the list below (exact copy).\par
- subtags: list of 2 to 5 strings in lower-case-kebab-case (no spaces, no uppercase, separated by hyphens) describing specific subareas / themes / techniques (e.g., 'llm', 'bert', 'graph-neural-networks', 'portuguese-language', 'sentiment-analysis', 'medical-imaging', 'time-series', 'covid-19', 'multi-objective', 'transfer-learning').\par
\smallskip
List of valid areas (choose ONE, copying the exact name):\par
- Machine Learning\par
- Deep Learning\par
- Computer Vision\par
- Natural Language Processing\par
- Reinforcement Learning\par
- Optimization / Metaheuristics\par
- Multi-Agent Systems\par
- Knowledge Representation / Reasoning\par
- Data Mining\par
- Recommender Systems\par
- Bioinformatics / Healthcare AI\par
- Robotics\par
- AI Ethics / Fairness / Explainability\par
- Other Applications\par
\smallskip
Rules:\par
1. Respond ONLY with the JSON, no markdown, no triple backticks, no comments.\par
2. If the paper applies deep learning to computer vision, prefer 'Computer Vision'. If it applies deep learning to NLP, prefer 'Natural Language Processing'. Use 'Deep Learning' only when the focus is a generic DL technique or architecture that does not fall into CV / NLP / RL.\par
3. Use 'Machine Learning' for traditional supervised or unsupervised methods (SVM, random forest, clustering, etc.) that are NOT deep learning.\par
4. If the focus is an application in healthcare / biology / medicine, use 'Bioinformatics / Healthcare AI'.\par
5. Subtags should be specific and short; avoid generic ones such as 'machine-learning' or 'deep-learning'.\par
\end{minipage}}
\caption{System prompt used by the topic-area classifier (sabiazinho-4, temperature 0), English translation. The production prompt lives in \texttt{scripts/classify\_topics.py}.}
\label{fig:prompt-topic}
\end{figure}

\paragraph{Contribution-type classification.} The classifier receives the paper title and the first 10{,}000 characters of the full text and returns a multi-label tag list of 1 to 3 entries drawn from an 4-class pool. The system prompt is shown in Figure~\ref{fig:prompt-contribution}.

\begin{figure}[H]
\centering
\fboxsep=0.7em
\fbox{\begin{minipage}{\dimexpr0.97\linewidth-2\fboxsep-2\fboxrule\relax}
\small\ttfamily\raggedright
You classify the TYPE of contribution of a scientific paper (not the subject, but rather WHAT KIND of scientific artifact it is).\par
\smallskip
Possible types:\par
- model: proposes or trains a new model\par
- algorithm: proposes a new algorithm or optimization technique\par
- survey: review or survey of existing methods\par
- empirical-study: systematic comparison of methods without proposing novelty\par
\smallskip
Multi-label allowed (1 to 3 types).\par
Respond ONLY with JSON: \{"types": [...]\}\par
\end{minipage}}
\caption{System prompt used by the contribution-type classifier (sabiazinho-4, temperature 0), English translation.}
\label{fig:prompt-contribution}
\end{figure}

\end{document}